\documentclass{aa}
\usepackage{graphicx}
\begin{document}

\title{ Sub-milliarcsec-scale structure of the gravitational
        lens B1600+434 }


\author{Alok R.~Patnaik
                  \inst{1}
                  \and
                  Athol J.~Kemball\inst{2}\\
               }

\offprints{A.~J.~Kemball}

\institute{
  Max-Planck-Institut f{\"u}r Radioastronomie,
  Auf dem H{\"u}gel 69, D-53121 Bonn, Germany \\
  \email{apatnaik@mpifr-bonn.mpg.de}
\and
  National Radio Astronomy Observatory, PO Box 0, Socorro, NM 87801,
  USA \\
  \email{akemball@nrao.edu}
               }

\date{Received date / Accepted date}

\abstract{In the gravitational lens system B1600+434 the brighter
  image, A, is known to show rapid variability which is not detected
  in the weaker image, B (Koopmans \& de Bruyn 2000).  Since
  correlated variability is one of the fundamental properties of
  gravitational lensing, it has been proposed that image A is
  microlensed by stars in the halo of the lensing galaxy (Koopmans \&
  de Bruyn 2000).  We present VLBA observations of B1600+434 at
  15~GHz with a resolution of 0.5~milliarcsec to determine the
  source structure at high spatial resolution.
  The surface brightness of the images are significantly different,
  with image A being more compact. This is in apparent contradiction
  with the required property of gravitational lensing 
  that surface brightness be preserved.  Our results suggest that
  both the lensed images may show two-sided elongation at this
  resolution, a morphology which does not necessarily favour superluminal
  motion. Instead these data may suggest that image B is
  scatter-broadened at the lens so that its size is larger than that
  of A, and hence scintillates less than image A.
  }

\maketitle

\section{Introduction}

The gravitational lens B1600+434 was discovered by Jackson et~al.
(1995). The radio source is a quasar at $z = 1.59$ and has two images
separated by 1.4~arcsec, which are denoted A and B. The main lensing
galaxy is identified as an edge-on spiral at $z = 0.41$, with an
additional lensing contribution from a nearby companion galaxy
(Jaunsen \& Hjorth 1997, Fassnacht \& Cohen 1998, Koopmans, de Bruyn
\& Jackson 1998). The background quasar is variable in both optical
and radio bands, thus allowing time delay measurements from monitoring
observations (Jaunsen \& Hjorth 1997, Koopmans et al. 2000). A time
delay of $47^{+12}_{-9}$~days has been determined from the radio
monitoring (Koopmans et al. 2000), and a delay of $51\pm4$~days from
the optical observations (Burud et al. 2000).

The radio observations of Koopmans et al. (2000) showed that the
brighter image, A, varied more rapidly than the weaker image, B. The
rapid variability in A was not found to be correlated with
corresponding rapid variability in B, although differential
variability between A and B on longer time scales allowed the
measurement of the time delay mentioned above. Correlated variability
is one of the fundamental characteristics of gravitational
lensing. Additional variability in A can only be explained by invoking
propagation effects which affect image A only (Koopmans \& de Bruyn
2000; hereinafter KdB). Such effects include a combination of
scattering in the lensing galaxy and scintillation in our own galaxy,
or microlensing in the halo of the lensing galaxy. KdB considered the
possibility that B might be scatter-broadened in the lens and thereby
would scintillate less than A. They argued however that the time scale
and the frequency dependence of the variability, taken in conjunction
with the high scattering measure required in the lens galaxy, did not
favour scintillation as a possible cause.  Instead, they proposed
microlensing of image A in the halo of the lensing galaxy as a preferred
alternative if the radio source has a superluminal component.

One of the ways to test these scenarios is to image the structure in
the lensed images at high spatial resolution, and to apply the
constraints imposed by gravitational lensing. In particular, the
surface brightness of gravitationally lensed images must be equal and
any significant departure would suggest that one of the images has
been affected by scattering more than the other. Since the
microlensing hypothesis requires the background quasar be a
superluminal source, the images, therefore, might be expected to show
evidence of this motion in their morphology, such as a core-jet
structure typical of such quasars. In addition, the polarization
properties of the images can differ if one of the ray paths is
affected either by scattering or microlensing.

With the above aims in mind, we observed B1600+434 at 15~GHz using the
VLBA with a resolution of 0.5~milliarcsec (mas), to detemine the
fine-scale structure and to measure image sizes for A and B. The
observations are described in Section 2 and the results are discussed
in Section 3.

\section{Observations and data analysis}

We observed B1600+434 on 2000 September 9/10 (1900 UT to 0500 UT) at
15~GHz using all ten telescopes of the VLBA. The data were recorded in
eight 8-MHz channels in dual circular polarization, yielding a total
recorded bandwidth of 32-MHz per polarization. Two-bit sampling
quantization was used. A single correlation centre was selected at
$\alpha=16^h01^m40.47^s,\ \delta=43^d16^m47.2^s$ (J2000). Since the
image separation of 1.4~arcsec is large compared with the resolution
of 0.5~mas, it is necessary to sample the data finely both in
frequency and time to avoid smearing of the visibility function and
the associated image distortions in the field of view.  The data were
therefore sampled at 0.5~MHz in frequency and 0.5~sec in time.

We observed 3C345 (J1640+3946), 1611+343 (J1613+3412) and 1749+096
(J1751+0939) as calibrators. The total integration time on the target
source B1600+434 was 195 min. We used 1749+096 to calibrate the
instrumental polarization, as well as to determine the bandpass
function.

The method of data analysis follows closely that described by Patnaik
et al. (1999) as used for the gravitational lens B1422+231.  The data
were analysed using the software package {\sc AIPS}, maintained by the
National Radio Astronomy Observatory (NRAO). After correcting for the
parallactic angle variation of the telescopes, amplitudes were
calibrated using the system temperatures and telescope gain factors as
a function of elevation.  The phase slopes and offsets across each
8-MHz band were aligned using the pulse-cal information from each
telescope. Standard fringe-fitting was performed on the
calibrators. After careful editing, the bandpass function was
determined using the source 1749+096. We followed the usual procedure
for calibrating instrumental polarization, using 1749+096 as the
polarization calibrator. Typical instrumental polarization amplitude
was measured to be $\sim$1.6\%.  Since B1600+434 has two images with
similar flux densities at a large angular separation, we had to
iterate the fringe-fitting process several times, each time using the
model of the source determined from the previous solution.

The final map was made with two sub-fields, centred on each lensed
image. We also made polarization maps of the source in Stokes $I,Q$
and $U$.

\section{Results and Discussion}

The maps of the lensed images, made using uniform weighting, are shown
in Fig.1. The maps have been restored using a circular gaussian with a
FWHM of 0.5~mas. We did not detect any polarized emission from either
component A or B, yielding a 5$\sigma$ limit to the percentage linear
polarization of 2.6\% for image A, and 3.3\% for image B. The rms
noise in the polarization map is 0.09~mJy/beam.

Table 1 lists the results of fitting a single gaussian component to
each of the images.  Two clear results emerge from these observations.
First, both the images are resolved, and A is more compact than B.
Secondly, the structure is elongated on either side of the peak in
both of the images. In order to quantify the errors of the size
measurements we divided the data into two sets by removing alternate
scans so as to preserve similar uv-coverage. The resulting maps were
then fitted with a single gaussian. The results show that the errors
for image A are $\sim0.01$~mas in each axis. However, the errors are
larger for image B by about 50\%. Note that the error estimates
derived in this way fold in most systematic calibration or imaging
errors which may affect the component size measurement, excluding
deviation of the component from non-Gaussian shape. A weak extension
is seen to the south of image B which appears to be genuine as it is
present in both the datasets mentioned earlier. We are not entirely
certain if it is real as there is no counterpart in image A. This can
be confirmed by another independent dataset. The measured size of
image B, however, is not affected by including this feature.

\begin{table}
\begin{center}
\begin{tabular}{llll}

Image    & Peak              & Total              & Size \\
         & in mJy/beam       & in mJy             & mas $\times$ mas, PA  \\
         &                   &                    &              \\
A        & 9.0$\pm$0.15      & 17.2$\pm$0.5       &
           0.69~$\times$~0.25, 103$^{\circ}$ \\
B        & 6.6$\pm$0.15      & 13.5$\pm$0.5       &
           0.73~$\times$~0.30, 93$^{\circ}$ \\
         &                   &                    &   \\

\end{tabular}
\end{center}
\caption{
 Peak (in mJy/beam) and total (in mJy) flux densities of the lensed
 images of B1600+434 at 15~GHz. 
 The image deconvolved sizes were obtained by fitting a single
 gaussian in the program {\sc JMFIT}. The error in the measured size
 is 0.01~mas in each axis for image A and about 0.015~mas for image B, 
 and 1$^{\circ}$ in position angle.
 The image separation of B from A is
 723.08$\pm$0.05~mas to the east (in RA) and 1189.16$\pm$0.05~mas to 
 the south (in Dec).
 }
\end{table}

The implication of the first result is that the surface brightnesses
of the two images differ significantly. The flux density ratio of A/B
is 1.27$\pm$0.05 whereas the ratio of image sizes is 0.80$\pm$0.15.
Secondly, the image morphology as measured is in apparent conflict
with a typical superluminal radio source in the sense that such radio
sources are core-dominated and must have one-sided jets due to
relativistic motion Doppler boosting of the knot components.

\begin{figure*}
\raisebox{-3.2cm}
{\begin{minipage}{0.3cm}
\mbox{}
\parbox{0.3cm}{}
\end{minipage}}
\begin{minipage}{8cm}
\mbox{}
\includegraphics[width=8cm]{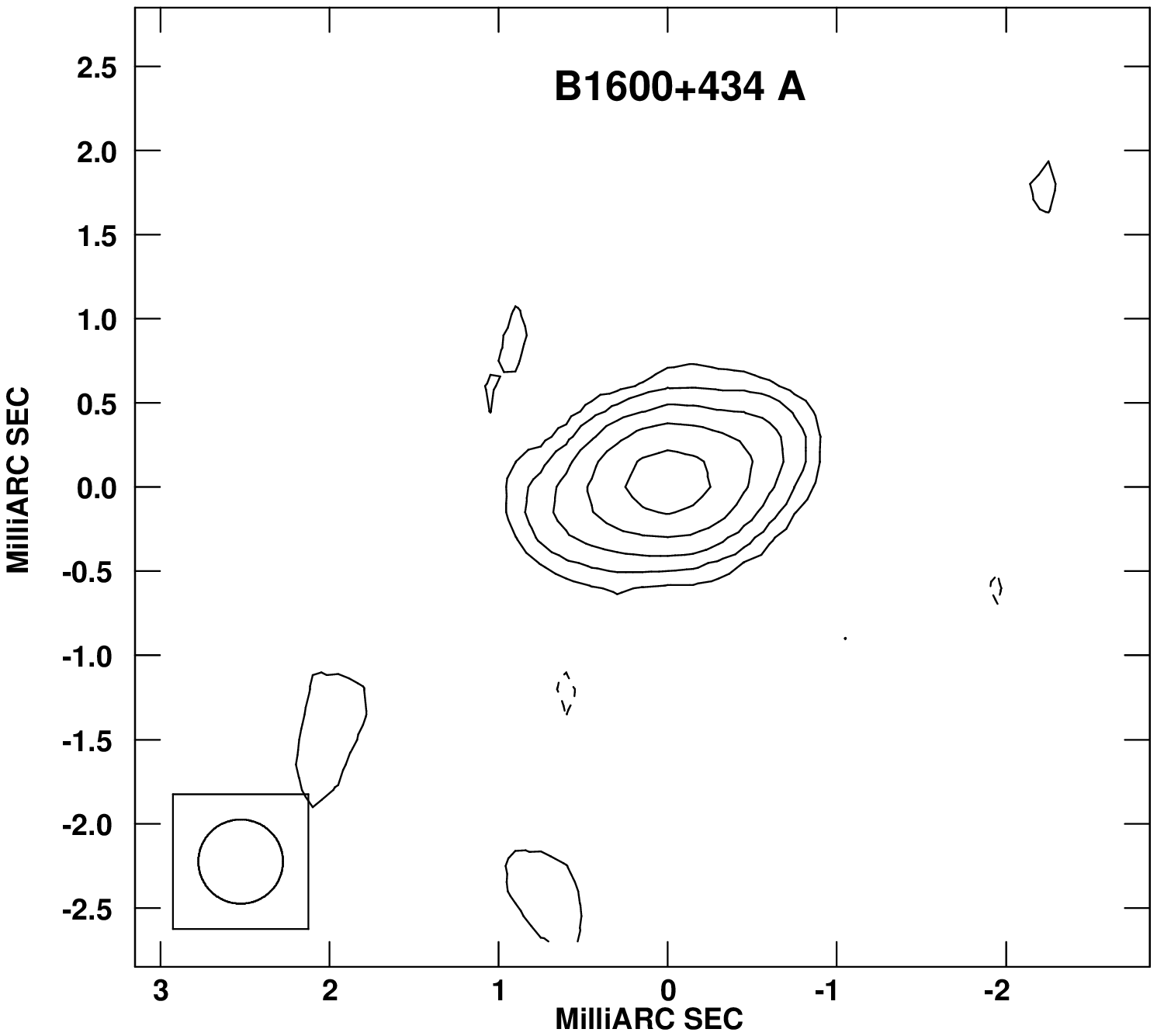}
\centering
\end{minipage}
\hspace{0.5cm}
\raisebox{-3.2cm}
{\begin{minipage}{0.3cm}
\mbox{}
\parbox{0.3cm}{}
\end{minipage}}
\begin{minipage}{8cm}
\mbox{}
\includegraphics[width=8cm]{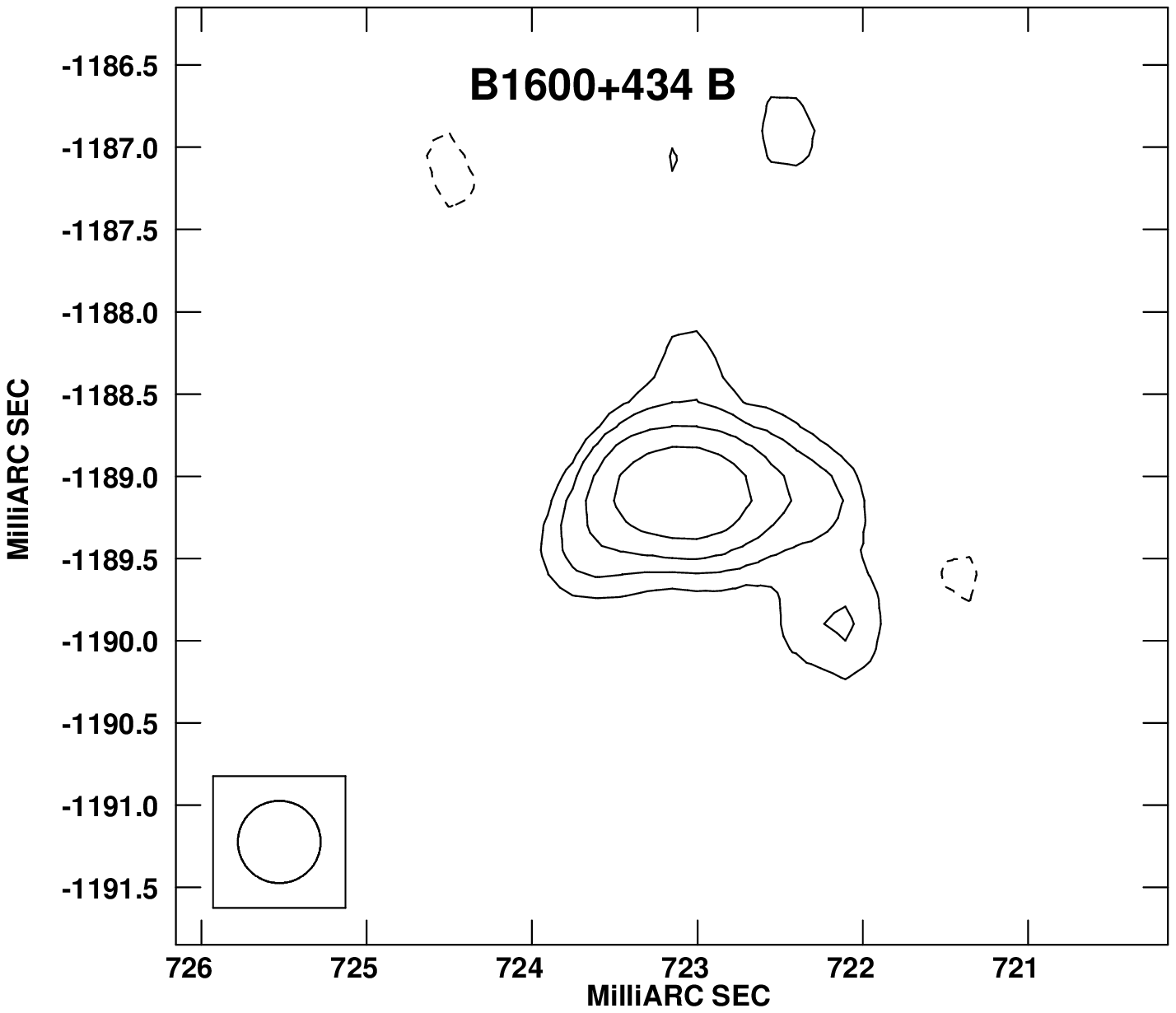}
\centering
\end{minipage}
\vspace{0.5cm}
\caption{ 
 15~GHz map of B1600+434 A (left) and B (right). The contour levels
 for both the images are $-$2, $-$1, 1, 2, 4, 8, 16, 32, 64,
  128 $\times$ the contour interval which is chosen as the 3$\sigma$
  noise in the map. For image A, contour interval is
  0.42~mJy~beam$^{-1}$ and peak flux density is 9.5~mJy~beam$^{-1}$
  and for image B contour interval is 0.42~mJy~beam$^{-1}$ and the
  peak is 6.7~mJy~beam$^{-1}$. The convolving beam of 0.5~mas circular
  gaussian is drawn at the lower left-hand corner in each map.
    }
\vspace{0.5cm}
\end{figure*}

In order to understand the higher modulation index (fractional rms
variability) of image A (2.8\% at 8.4~GHz), KdB considered the
possibility that the size of B (modulation index of 1.6\% at 8.4~GHz)
might be larger through scatter-broadening by the ionised interstellar
medium in the lens galaxy. The fact that the surface brightnesses do
not agree, and in particular, that the weaker image B is more
extended, suggests that B has been affected by propagation through the
lensing galaxy more than A.  This would support the idea that B is
somewhat broadened by scattering in the lensing galaxy. We note that
its ray path passes close to the bulge of this edge-on spiral galaxy
whereas the ray path of A goes through the halo of the lens. For the
two images to scintillate in our Galaxy, the required sizes are 62
$\mu$arcsec and 108 $\mu$arcsec for A and B respectively (KdB). Our
observations cannot resolve such sizes and our measured sizes are
considerably larger than these.  We note, however, that the observed
images have been parametrised as single gaussian components. This does
not exclude any sub-structure in the source. Therefore, our result
that image B is larger in size than image A is in general agreement
with the expectation from scintillation taking place in our Galaxy.

KdB measured a modulation index $\le$1.2\% at 1.4~GHz and 3.7\% at
5~GHz. If the variability was due to scintillation, then the
modulation index should have increased going to lower frequencies.
However, if there is frequency-dependent structure then the modulation
index can be smaller at lower frequencies. A flat spectrum source
likely consists of more than one self-absorbed component (Cotton et
al.  1980). It is then possible that the component which is compact
and scintillates at 5.0~GHz, which may for example be the core, does
not contribute much to the flux density at 1.4~GHz due to
self-absorption.  Some other component in the jet might be brighter at
1.4~GHz but this component would be more extended and therefore
unlikely to scintillate. Therefore, a smaller modulation index at a
lower frequency does not necessarily rule out scintillation occurring
at a higher frequency. Such frequency-dependent structure has been
noted in the gravitational lens B0218+357 (Patnaik \& Porcas 1999).

The alternative explanation proposed by KdB is that A is microlensed
in the halo of the lensing galaxy. However, for the microlensing to
match the time-scale of variability, the model used by KdB requires
that the background source be a superluminal source with a knot the
size of a few $\mu$arcsec containing $\sim$ 5-11\% of the flux
density, moving at an apparent speed in the range $\sim$ 9-26c.  The
morphology of a superluminal source will of necessity be that of a
bright core and one-sided jet, due to Doppler boosting of a
relativistically moving knots in the jet. The detection of extensions
on either side of the peak in both the lensed images here is contrary
to the above expectation. It would, therefore, appear that
superluminal motion in the background source is not favoured and hence
the microlensing caustic crossing time will be much longer than the
observed variability. However, it is also possible that the peak we
observe is not necessarily the core; it might be a bright knot and
thus still allow the possibility that the source could be
intrinsically one-sided. We do not have sufficient resolution to
distinguish this level of detailed structure.  If we assume that the
peak corresponds to the core, which may be a reasonable assumption
given the fact that these observations are at 15~GHz and with 0.5~mas
resolution where cores usually brighter than knots, then it would
impliy that microlensing combined with superluminal motion in the
background source may not be a viable explanation for the rapid
variability of the lensed image A.

As an example, the two brightest images in the lens system B1422+231
have elongation produced by lensing (Patnaik \& Porcas 1998, Patnaik
et al. 1999). In this lens system the background radio source is also
a flat spectrum object and at milliarcsec-scales does not have the
standard core-jet structure. It has been suggested that the magnification
in this source might be very high and that the background object
could possibly be a radio-quiet object. A similar scenario is possible for
B1600+434. Certainly, in both B1422+231 and B1600+434, no jet has been
detected and both sources show two-sided emission.  If the extension
of the images we detect is due to stretching of the extended
background emission by gravitational lensing, then we cannot be sure
of the direction of the jet.  However, should there be superluminal
motion (which can be as much as 0.15~milliarcsec per year for the
bright quasars), we would detect changes in image size after a year
or two. 

\section{Conclusions}

We have observed the gravitational lens system B1600+434 at 15~GHz
using the VLBA at a resolution of 0.5~milliarcsec. We find that the
two lensed images are resolved and have elongation on either side of
the peak, and that the surface brightnesses of the images are
significantly different. The image shapes do not appear to favour
superluminal motion in the background radio source and hence the
interpretation that A is microlensed is not favoured by these
observations at this epoch.  Our results are consistent with the view
that image B is perhaps scatter-broadened by the interstellar medium
of the lens galaxy, and therefore scintillates less than image A in
our Galaxy.  However, VLBI observations of the source at a range of
other observing frequencies are required to provide further insight
into this question.

\begin{acknowledgements}
  
  We thank Peter Schneider for helpful discussion and J. Schmid-Burgk
  and Karl Menten for a careful reading of the manuscript. We would
  like to thank the referee A.G. de Bruyn for critical comments and
  helpful advice. The National Radio Astronomy Observatory is a
  facility of the National Science Foundation operated under
  cooperative agreement by Associated Universities, Inc.

\end{acknowledgements}

\end{document}